\documentclass{aa}  

\usepackage{graphicx}
\usepackage{txfonts}

\usepackage[colorlinks=true,
    linkcolor=blue,
    citecolor=blue]{hyperref}

\usepackage{amssymb}
\usepackage{amsmath}
\usepackage{wasysym}
\usepackage{float} 
\usepackage{wrapfig}
\usepackage{physics}
\usepackage{array}
\usepackage{comment}
\usepackage{orcidlink}

\newcommand\bb[1]{\mbox{\boldmath{$#1$}}}
\renewcommand\grad{\bb{\nabla}}
\newcommand\bcdot{\,\bb{\cdot}\,}

\newcommand\btimes{\,\bb{\times}\,}
\newcommand\rmd{\mathrm{d}}

\usepackage{soul} 
\usepackage{xcolor}

\begin{document}

   \title{Non-thermal electron acceleration in turbulent, post-flare coronal loops }

   \author{C. Mora\thanks{clarissa.mora@kuleuven.be}\inst{\ref{inst1}}\orcidlink{0009-0003-9899-2182}, F. Bacchini\inst{\ref{inst1},\ref{inst2}}\orcidlink{0000-0002-7526-8154}, \and R. Keppens\inst{\ref{inst1}}\orcidlink{0000-0003-3544-2733} }
   
   \institute{Centre for mathematical Plasma Astrophysics, Department of Mathematics, KU Leuven, Celestijnenlaan 200B, B-3001 Leuven, Belgium \label{inst1}
   \and Royal Belgian Institute for Space Aeronomy, Solar-Terrestrial Centre of Excellence, Ringlaan 3, B-1180 Uccle, Belgium \label{inst2}}

   \date{Received ; accepted }

 
  \abstract
  {The generation of energetic, non-thermal electrons during solar flares plays a critical role in energy transportation from the corona to the chromosphere, producing regions of observed intense X-ray emission. Turbulence in post-flare loops, particularly from Kelvin--Helmholtz instabilities (KHIs), has been suggested and investigated as a mechanism for trapping and accelerating electrons in such scenarios. }
  {Starting from past results, we aim to characterise the energisation process of electrons trapped in a turbulent post-flare looptop, quantifying the contributions of different acceleration mechanisms, and establishing a coherent numerical framework for describing particle energetics.}
  {We performed test-particle simulations with the guiding-centre approximation in addition to a 2.5D magnetohydrodynamic model of a time-evolving, post-flare coronal looptop. We implemented an improved formulation of the guiding-centre equations, which explicitly conserves energy, enabling a consistent analysis of electron acceleration in the turbulent plasma.}
  {We find that, in the plasma turbulence inside the looptop, electrons develop supra-thermal energy distributions with tails compatible with hard X-ray emission. The dominant energisation channel arises from perpendicular gradient effects in the form of second-order, Fermi-like stochastic acceleration, while curvature effects are dominant for particles on long trajectories. Statistical correlations with the measured particle pitch angle confirm that the strongest acceleration occurs for electrons trapped in bouncing motions within turbulent magnetic structures.}
  {Our results provide an understanding of how KHI-induced turbulence in coronal looptops produces and sustains populations of trapped non-thermal electrons, supporting the interpretation of observed X-ray sources. We dissect and clarify the relative roles of different magnetic effects and the emergence of stochastic, Fermi-like energisation. We also demonstrate the advantages of the improved guiding-centre-approximation (GCA) formalism on a simple reproducible test, for the future benchmarking of GCA implementations.}

\keywords{acceleration of particles – Sun: corona – Sun: flares – Sun: X-rays – turbulence}
\titlerunning{Electron energisation in turbulent coronal loop plasma}
\maketitle
\section{Introduction}
The generation of energetic electrons is a hot topic in the study of solar flares, as particle acceleration and the ensuing collisional energy loss act as an important path of energy transfer and transport (\citealt{kerr2020}). During a solar-flare event, up to $10^{32}$~erg of energy is released via magnetic reconnection, and up to 50\% of the released energy is involved in the generation of non-thermal electrons (\citealt{Aschwanden_2017}). Potential acceleration mechanisms include electric DC-field acceleration, stochastic acceleration, and shock acceleration \citep{Aschwanden2005}. 

The estimation of total energy and number of non-thermal electrons in these sources has proven to be quite challenging, partly due to the lack of defined observational constraints of the acceleration region and the presence of a strong, bright thermal emission that covers the non-thermal spectral features, especially in the low-energy range. For example, of considerable interest to understanding of the nature of particle acceleration to high energies are the ratios of the number densities of nonthermal and thermal electrons ($n_{nth}$ and $n_{th}$, respectively) to the total number density of background electrons $n_p = n_{th}+n_{nth}$. The instantaneous fraction of accelerated electrons in the flaring corona, $n_{nth}/n_p$, is discussed in both past and more recent literature (\citealt{Kruker2010,Oka_2013,KrukerBattaglia2014,Fleishman2022}, \citealt{Kontar2023}), but results and methods differ, with quite variable values reported across the $0.01 \leq n_{nth}/n_p \leq 1 $ range; thus, there is room for further investigation. Although we focused here on electron acceleration, it is worth noting that proton precipitation is also expected to occur, and a recent re-analysis of RHESSI's hard X-ray and $\gamma$ -ray findings, combined with white-light observations for an extreme X25 flare in 2003, finds both electron and proton precipitation sites to coincide with the white-light flare ribbons \citep{krucker2025}. Alternative relevant mechanisms of energy transport in flares, which may operate alongside transport by non-thermal electrons and ions, include thermal conduction following direct heating of the corona, and coronally generated Alfvénic waves damped in the lower atmosphere (\citealt{kerr2020}; see \citealt{Kerr2023} for a review).\\
Energetic non-thermal electrons are thought to be generated at coronal height after a reconnection event. They most likely produce a first generation of radiation in the hard X-ray (HXR), slightly above the looptop, as first noted by \citealt{Masuda1994} and observed in several events afterward (\citealt{Petrosian2002}, \citealt{KrukerBattaglia2014}). Subsequently, they move to the chromosphere along magnetic-field lines and produce footpoint HXR sources, losing most of their energy there through collisions (\citealt{Kruker2008}; \citealt{Benz2017}). This intense chromospheric heating drives chromospheric evaporation, through which flare loops are filled with upflowing heated plasma and become bright in soft X-rays (SXRs) \citep{Milligan_2009}. Indirect proof of this phenomenon is given by the observational finding known as the Neupert effect \citep{Neupert1968}: the temporal profile of the HXR flux is closely related to the time derivative of the SXR profile, for which the time integral of the non-thermal emissions tracks the time profile of the thermal emissions. It has been shown that the flux of energetic electrons in the looptop can be several times (a factor of 1.7-8) higher than the flux of electrons reaching the footpoints during the impulsive phase of the flare (\citealt{simoens}), suggesting an accumulation of electrons in the looptop. Thus, electrons must be subjected to a trapping mechanism that holds a significant fraction of them inside the coronal loops where they potentially undergo a second stage of acceleration. This creates a secondary generation of radiation from the looptop that has been relatively unexplored, particularly in terms of the associated dynamics of energetic particles emitting in the X-ray wavelengths.\\
\citealt{Fang_2016} proposed a scenario for the generation of non-thermal electrons inside the looptop, in which evaporation from the chromosphere produces turbulence at coronal height via the Kelvin--Helmholtz instability (KHI); electrons could be accelerated in this turbulence through stochastic processes, providing means of origin for the second generation of radiation mentioned above. The KHI is due to the interaction of evaporation upflows from both post-flare loop footpoints, which is almost surely asymmetric and hence quite liable to shear-flow-driven deformations \citep{Ruan2018}. In such a scenario, \citealt{Bacchini2024} used test particle methods to show how KHI-driven turbulence can confine accelerated electrons inside the looptop long enough to create the strong looptop HXR sources that were observed. In that work, they analysed the contribution of KHI turbulence to the production and trapping of solar-flare non-thermal electrons inside coronal looptops, but the precise mechanism of particle energisation was not yet fully documented. Recently, several 3D magnetohydrodynamic (MHD) simulations clarified that turbulent looptops are omnipresent (\citealt{Shen2022,Wang2023,Russell2024}) and may result from KHI \citep{Ruan_2023} during the impulsive-phase or Rayleigh-Taylor processes throughout the gradual phase \citep{Ruan2024}; thus, it is of interest to revisit how turbulent looptops energise and trap particles in detail. \\
To investigate the electron acceleration in a large-scale phenomenon such as a flare, test-particle methods can be employed, owing to their relatively reduced costs compared to fully kinetic simulations. Test particles evolved on top of an MHD simulation do not provide feedback to the MHD fields, lacking self-consistent kinetic mechanisms. A step towards self-consistent treatments where analytic electron beam models are integrated in multi-dimensional MHD studies was realised in \citet{Ruan2020} and \citet{Druett2023,Druett2024}, but these do not address the actual energisation process of the beams. Test-particle approaches have been used to qualitatively study particle motion and acceleration and in simulations of coronal flares, even those as large as the entire flare
loop (e.g. \citealt{Gordovskyy_2010,Gordovskyy_2014} ; \citealt{Zhou_2015}, \citealt{Zhou_2016}; \citealt{Gordovskyy_2020}; \citealt{Kong2022}). Alternative relevant methods to combine multi-scale dynamics of particle acceleration and 3D MHD setups more self-consistently can be found in, for example, \citealt{Baumann2013} and \citealt{Li2025}.
Recent studies of electron acceleration in erupting flux-rope settings also point to the need to resolve mini flux ropes in the current sheet underneath, where their helicity determines the electron impact sites \citep{Wu2025}.

In this work, we revisited the test-particle simulations presented in \citealt{Bacchini2024} to focus specifically on the energisation process of particles trapped within the turbulent coronal loop. To do so, we employed the state-of-the-art MHD simulations of looptop turbulence first presented in \citealt{Ruan2018} and analysed the energy dynamics and evolution of a distribution of test electrons inside the looptop. The test particles evolve in an MHD background, for which we employed a simplified model of chromospheric evaporation in a post-flare loop state. This model is agnostic to the primary reconnection process above the looptop, which we did not model. We simply assumed that footpoint heating occurs, and let fluid flows meet and mix in the looptop region. Our main aim is to quantify precisely how charged particles energise within a pre-existing turbulent state.\\
Our first objective is to define the most suitable numerical approach to modelling test-particle motion that guarantees accurate representation and analysis of particle energetics. We also aim to investigate (i) the highest energies up to which electrons can be accelerated to produce the strong looptop HXR sources that have been observed, (ii) the stages through which energisation proceeds (e.g. in different directions) and magnetic energy is converted into thermal-kinetic energy of particles, and (iii) the leading acceleration mechanism in the turbulent structure of the looptop plasma. \\
This paper is organised as follows. In Sect.~\ref{sec:numerical}, we review the MHD and test-particle models employed in our simulations; in Sect.~\ref{sec:sec3}, we investigate aspects of energy conservation in our numerical approach; in Sect.~\ref{sec:Elec}, we quantify and characterise particle acceleration; and, finally, in Sect.~\ref{conclusions}, we summarise our results.

\section{Numerical method}\label{sec:numerical}
\subsection{Ideal MHD setup}\label{sec:MHD}
Our simulations were performed with the open-source MPI-AMRVAC 3.0 code (\citealt{keppens2023}). We ran an ideal-MHD simulation (i.e. without including resistivity or other diffusion terms) with a simplified, 2.5D (i.e., with 2 spatial dimensions and 3 vector components) flare model, in which the reconnection current sheet is not included. The full setup is explained in detail in \citealt{Ruan2018}, and we briefly summarize it here. The corona and the chromosphere are included at the bottom boundary of our setup, for which we employed a numerical domain of $-40~\mathrm{Mm} \leq x \leq 40~\mathrm{Mm}$ and $0~\mathrm{Mm} \leq y \leq 50~\mathrm{Mm}$. The base numerical resolution consists of 128 $\times$ 80 cells; via adaptive mesh refinement (AMR), we attain an effective resolution of 2048 $\times$ 1280 cells at the highest AMR level (using 5 levels). This implies an effective resolution of about $40~\mathrm{km}$, which is comparable with observational limits. In addition, while performing time-evolving test-particle simulations, we allowed the resolution to increase even further by setting the maximum AMR level to seven. As a result, inside the turbulent looptop we now achieve an effective resolution of 8192 $\times$ 5120 cells; i.e. our minimum grid spacing is $\sim 10~\mathrm{km}$. 

At the beginning of the MHD run, an arcade magnetic field in the $xy$ plane was set up with a guide field component along $z$. Localised heating was added at the chromospheric footpoints of selected loops to produce evaporation flows. These flows generate a post-flare loop by filling the heated loop with hot, dense plasma. Before evaporation occurs, the magnetic-field strength at the footpoints is $80~\mathrm{G}$, and at the looptop it is $50~\mathrm{G}$. The footpoint heating rate is given by Eqs.~(12)--(16) of \citealt{Ruan2018}; the resulting simulation has parameters $c_1=1.288 \times 10^{13}~\mathrm{erg}~\mathrm{cm}^{-2}$. The parameter $\lambda_t$ in Eq.~(15) of \citealt{Ruan2018} is set to $60~\mathrm{s,} $ and other parameters are provided in Sect.~2 of their paper. These parameters are inspired by energy estimates from observations: assuming that the depth of the loop in the out-of-plane direction is $\sim 10~\mathrm{Mm}$, the energy injected into the chromosphere is $\sim 10^{30}~\mathrm{erg}$, which is within the range of M-class flares (\citealt{aschwaden2016}). These parameters also provide maximum energy-deposition rates of around $3 \times 10^{10}~\mathrm{erg}~\mathrm{cm}^{-2}~\mathrm{s}^{-1}$ at the footpoints, which align well with observational and modelling estimates (e.g. \citealt{Allred_2015} and references therein). The unit time in this work, $t_0 = 86~\mathrm{s}$, is the timescale on which acoustic waves travel a distance of $L_0 = 10~\mathrm{Mm}$ in a typical coronal-plasma environment with a temperature of $T_0 = 1~\mathrm{MK}$. This initial MHD setup was run for $2.5t_0$; at that point, we observed that the turbulence in the looptop was well developed, and the five-level MHD run was stopped before we injected particles. Several of the multi-dimensional MHD studies, with (anomalously prescribed) reconnection included, indeed demonstrate looptop turbulent regions within minutes of the reconnection onset (\citealt{Shen2020}, \citealt{Ruan_2023}).

\subsection{Particle motion: Guiding-centre approximation}
To study the dynamics of electrons in the looptop turbulence of our MHD simulations, we evolved ensembles of electrons using the test-particle module of MPI-AMRVAC 3.0 (\citealt{keppens2023}). These particle ensembles were tracked in the time-evolving MHD background described in Sect.~\ref{sec:MHD}, starting from the MHD state at $t = 2.5t_0$, when turbulence in the looptop is well developed. An ensemble of $\sim 10^6$ electrons was initialised according to a Maxwellian distribution with a temperature of $T = 20~\mathrm{MK}$ (approximately the average temperature inside the looptop; see \citealt{Bacchini2024}) since the average magnetic-field strength inside the looptop is $B \sim 5\mbox{--}40~\mathrm{G}$, and the maximum electron gyroradius is $\rho_C = \sqrt{m_ekT}/(|q_e|B) \sim 1~\mathrm{m}$ (where $m_e$ and $q_e$ are the electron mass and charge) at initialisation. Because our MHD system size is of the order of $10^7~\mathrm{m}$, spatially resolving the gyromotion on the numerical grid would require unachievable resolution; the same applies for the gyration timescales, which are much faster than the MHD dynamical time. For this reason, we chose to evolve our test particles according to the guiding-centre equations of motion, which are appropriate when the gyroradius is negligible in size and the gyrofrequency is very large.\\
The system of relativistic equations of motion under the guiding-centre approximation (GCA; e.g. \citealt{Vandervoort1960}; \citealt{Northrop1963}) evolves the guiding-centre position, $\bb{R}$; parallel four-velocity, $\bb{u}_\parallel$; and magnetic moment, $\mu = m u_\perp^2/(2B /\kappa) $ of a charged particle of mass, $m,$ and charge, $q,$ as
\begin{equation}
    \frac{\rmd\bb{R}}{\rmd t} = \frac{u_\parallel}{\gamma} \bb{b} + \bb{v}_E +  \bb{v}_{curv}+\bb{v}_{pol} + \bb{v}_{{\boldsymbol{\nabla}} B} + \bb{v}_{rel},\label{eq:gcadRdt}
\end{equation}
\begin{equation}
    \frac{\rmd u_\parallel}{\rmd t} = \frac{q}{m} E_\parallel + a_{curv} + a_{mirr},
    \label{eq:gcadupardt}
\end{equation}
\begin{equation}
    \frac{\rmd\mu}{\rmd t} = 0.
    \label{eq:gcadmudt}
\end{equation}
Here, the spatial part of the particle four-velocity $\bb{u} = \gamma \bb{v}$ (where $\gamma = 1/ \sqrt{1- v^2/c^2}= \sqrt{1+u^2/c^2}$ the Lorentz factor) is split into the parallel and perpendicular\footnote{Note that with $u_\perp$ we indicate the perpendicular particle velocity linked to the particles gyromotion; i.e. excluding the velocity-drift terms that determine the guiding-centre motion across magnetic-field lines. These terms are typically much smaller than $u_\parallel$, and hence we can safely approximate $u_\perp^2 \simeq u^2 - u_\parallel^2$. If needed, a better approximation is given by $u_\perp^2 \simeq u^2 - u_\parallel^2 - v_E^2 \gamma^2 $, since $v_E$ is the dominant drift term (e.g. \citealt{bacchini2020}).} components $u_\parallel = \bb{u} \bcdot \bb{b}$ and $u_\perp \simeq \sqrt{u^2 - u_\parallel^2}$ with respect to the magnetic field $\bb{B}$ with unit vector $\bb{b} = \bb{B} /B$ . The motion of the particles guiding centre  (Eq.~\eqref{eq:gcadRdt}) is described as a superposition of motions along and across magnetic-field lines, indicated by a number of `drift' velocity terms. The dominant term $\bb{v}_E = \bb{E} \btimes \bb{B} /B^2$ is the $E \times B$ drift, with the associated Lorentz factor $\kappa = 1 / \sqrt{1 - v^2_E /c^2}$. Similarly, the parallel momentum evolves Eq.~\eqref{eq:gcadupardt} according to the parallel-acceleration term $a_\parallel = qE_\parallel /m$ (where $\bb{E}_\parallel = \bb{E} \bcdot \bb{b} = 0$ in ideal MHD), as well as the curvature acceleration $a_{curv}$ and the mirror acceleration term $a_{mirr}$. Finally, the last equation (Eq.~\eqref{eq:gcadmudt}) adopts the usual adiabatic-invariant assumption, such that the magnetic moment remains constant in time. In all these expressions, terms proportional to time derivatives of the electromagnetic fields are ignored, under the assumption that particle dynamics takes place on much faster timescales than those over which MHD fields typically evolve (the slowly varying fields condition from adiabatic motion treatment).
Details on the explicit nature and role of drift motions and confining trajectories in the test-electrons dynamics were investigated in Sect.~2.2 and Sect.~3 of \citealt{Bacchini2024} and are assumed as valid throughout this work.\\

As is typical for the case of GCA equations, the employed integration method for particle motion is an explicit fourth-order Runge--Kutta (RK4) algorithm (e.g. \citealt{Ripperda_2018} and Sect.~7 of \citealt{keppens2023}). We note that to evaluate the various drifts and accelerations, it is necessary to properly interpolate electric and magnetic-field gradients from the high resolution MHD run, in both space and time. That must, for example, ensure that the ideal MHD setup always has exactly $E_\parallel=0$, since the electric field follows from the MHD fluid speed, $\bb{v}_\mathrm{MHD,}$ as $\bb{E}=-\bb{v}_\mathrm{MHD}\btimes \bb{B}$. In what follows, we emphasise a particular numerical shortcoming of the standard way of handling these GCA equations, which we identified and cured while investigating the acceleration processes at play in turbulent MHD fields. Since this is part of the new findings of this paper, we demonstrated and documented it on a simple and reproducible test for future reference, before applying it to the loop-top turbulent regime. Our main finding is a reformulation to ensure energy conservation for particles, which necessarily relaxes the adiabatic invariance assumption.

\section{Energy-conserving GCA particle integration}\label{sec:sec3}
In the following section, we address the first objective of this study. We find that, by solving Eqs.~\eqref{eq:gcadRdt}--\eqref{eq:gcadmudt} with a standard RK4 approach, a discrepancy arises between the expected evolution prescribed by GCA theory along the perpendicular direction and the actual dynamics of particles inside the simulation. We first conducted a number of critical tests with few particles and analytically prescribed electromagnetic fields to identify the aspects in the numerical treatment of the particles that could potentially lead to an energy inconsistency. Here, we only show the most relevant test case of a 2D Alfvén-wave propagation in a uniform magnetised medium in Sect.~\ref{sec:AW}.

We argue the need for a new numerical implementation of the magnetic-moment GCA equation \eqref{eq:gcadmudt}, and, in doing so, we developed a consistent framework for numerically exploring particle energetics (i.e. fulfilling energy conservation). This new energy-conserving GCA formalism was then applied to the realistic physical setup of test electrons in the coronal loop discussed in Sect.~\ref{sec:Elec}. We note that the original work \citep{Bacchini2024} actually used the standard method for test-particle studies based on adhering to Eq.~\eqref{eq:gcadmudt}, and we further clarify which aspect is improved and revisited in this paper.

\begin{figure}
 \centering
\includegraphics[width=8.5cm]{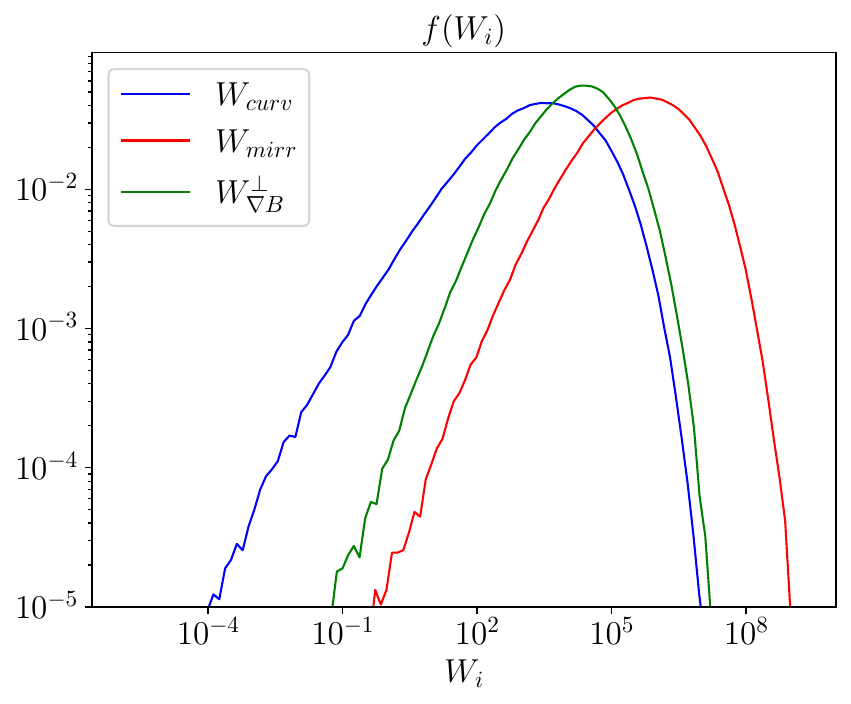}
\caption{Distribution functions for individual '\textit{i} ' type contributions to particle energisation at the initial test-particle simulation time, $t = 2.5t_0,$ inside the turbulent looptop.}\label{fig:dNdWi}
\end{figure}

\subsection{GCA energy equations}\label{sec:sec3_1}
Following the decomposed approach of the GCA description, one can define an energy equation for parallel and perpendicular particle motion, respectively. The parallel energy evolution can be obtained directly from Eq.~\eqref{eq:gcadupardt} for the relativistic momentum $u_\parallel=\gamma v_\parallel$, multiplied on both sides by a factor of $\gamma v_\parallel$:

$$
\begin{aligned}
(\gamma v_\parallel)\frac{\rmd (\gamma v_\parallel)}{\rmd t} = (\gamma v_\parallel) \biggl\{&  - \frac{\mu}{m\gamma} \bb{b} \bcdot \grad \biggr(\frac{B}{\kappa}\biggl) \\
 & + \bb{v}_{E} \bcdot [ \gamma v_{\parallel}(\bb{b} \bcdot \grad) \bb{b} + \gamma (\bb{v}_{E} \bcdot \grad) \bb{b}] \biggr\}.
\end{aligned}
$$

By simple algebra, one obtains
\begin{equation}
  \begin{aligned}
    \frac{1}{2}\frac{\rmd(\gamma v_\parallel)^2}{\rmd t} = & - \frac{\mu}{m} v_\parallel \bb{b} \bcdot \grad \biggr(\frac{B}{\kappa}\biggl)  
    \\
    & + \bb{v}_{E} \bcdot [ (\gamma v_{\parallel})^2(\bb{b} \bcdot \grad) \bb{b} + \gamma^2 v_{\parallel} (\bb{v}_{E} \bcdot \grad) \bb{b}]. \label{Wpar}
    \end{aligned}
\end{equation} 
An evolution equation for the perpendicular energy can be found from the magnetic-moment conservation condition \eqref{eq:gcadmudt} by explicitly substituting $\mu = m(\gamma v_\perp)^2/(2B / \kappa)$:
$$\begin{aligned}
    \frac{\rmd}{\rmd t} \biggl(\frac{m(\gamma v_\perp)^2}{2 B/\kappa}\biggr) = 0& \\
    \Rightarrow\frac{m}{2}\biggl[\frac{1}{ B/\kappa}\frac{\rmd(\gamma v_\perp)^2}{dt} \ - \ \frac{(\gamma v_\perp)^2}{(B/\kappa)^2}&\frac{\rmd}{\rmd t}\frac{B}{\kappa} \biggr] = 0 \ .
\end{aligned} $$
By using the total (Lagrangian) derivative definition in the GCA formalism $\rmd/\rmd t = \partial/\partial t \ + \ (v_\parallel \bb{b}+ \bb{v_E}) \bcdot \bb{\nabla}$ \citep[][]{Northrop1963}, one obtains
\begin{equation}
\centering
   \frac{1}{2} \frac{\rmd(\gamma v_\perp)^2}{\rmd t}  = \frac{(\gamma v_\perp)^2}{2B/\kappa} \left[  v_\parallel \bb{b} \bcdot \grad \left(\frac{B}{\kappa}\right) +  \bb{v}_{E} \bcdot \grad \left(\frac{B}{\kappa}\right) \right], \label{Wperp} 
\end{equation}
where the partial time derivative, $\partial B / \partial t \approx 0,$  has been neglected under the slowly varying fields approximation. Finally, from the sum of Eqs.~\eqref{Wpar} and \eqref{Wperp}, the total energy evolves as 
\begin{equation}
\begin{aligned}
    \frac{1}{2}\frac{\rmd(\gamma v)^2}{\rmd t} = \ &\frac{\mu}{m} \bb{v_{E}} \bcdot \grad \left(\frac{B}{\kappa}\right) \ +\\
    & + \bb{v}_{E} \bcdot \left[ (\gamma v_{\parallel})^2(\bb{b} \bcdot \grad) \bb{b} + \gamma^2 v_{\parallel} (\bb{v}_{E} \bcdot \nabla) \bb{b} \right]. 
    \end{aligned}
    \label{Wtot}
\end{equation} 
Here, we can distinguish three contributions to the particle energy evolution from the underlying MHD background. These are listed below.
\begin{itemize}
    \item[\textit{(i)}] Magnetic-mirror effects: 
    \begin{equation}
    W_{mirr} = \pm \frac{\mu}{m} v_\parallel \bb{b} \bcdot \grad \left(\frac{B}{\kappa}\right), \label{eq:Wmirr}
    \end{equation} 
    acting equally in both directions, with opposite signs. In the total energy equation~(\ref{Wtot}), these cancel out exactly. \\
    \item[\textit{(ii)}] Curvature effects: 
    \begin{equation}
        W_{curv}=\bb{v}_{E} \bcdot [ (\gamma v_{\parallel})^2(\bb{b} \bcdot \grad) \bb{b} + \gamma^2 v_{\parallel} (\bb{v}_{E} \bcdot \grad) \bb{b} ],
    \end{equation}
    only contributing to parallel-energy evolution.\\
    \item[\textit{(iii)}] Perpendicular-gradient effects:
    \begin{equation}
    W^\perp_{\boldsymbol{\nabla} B} = \frac{\mu}{m} \bb{v}_{E} \bcdot \grad \left(\frac{B}{\kappa}\right),
    \end{equation} 
    exclusive to perpendicular-energy evolution.
\end{itemize}

\begin{figure*}
\centering
  \includegraphics[width=0.88\textwidth]{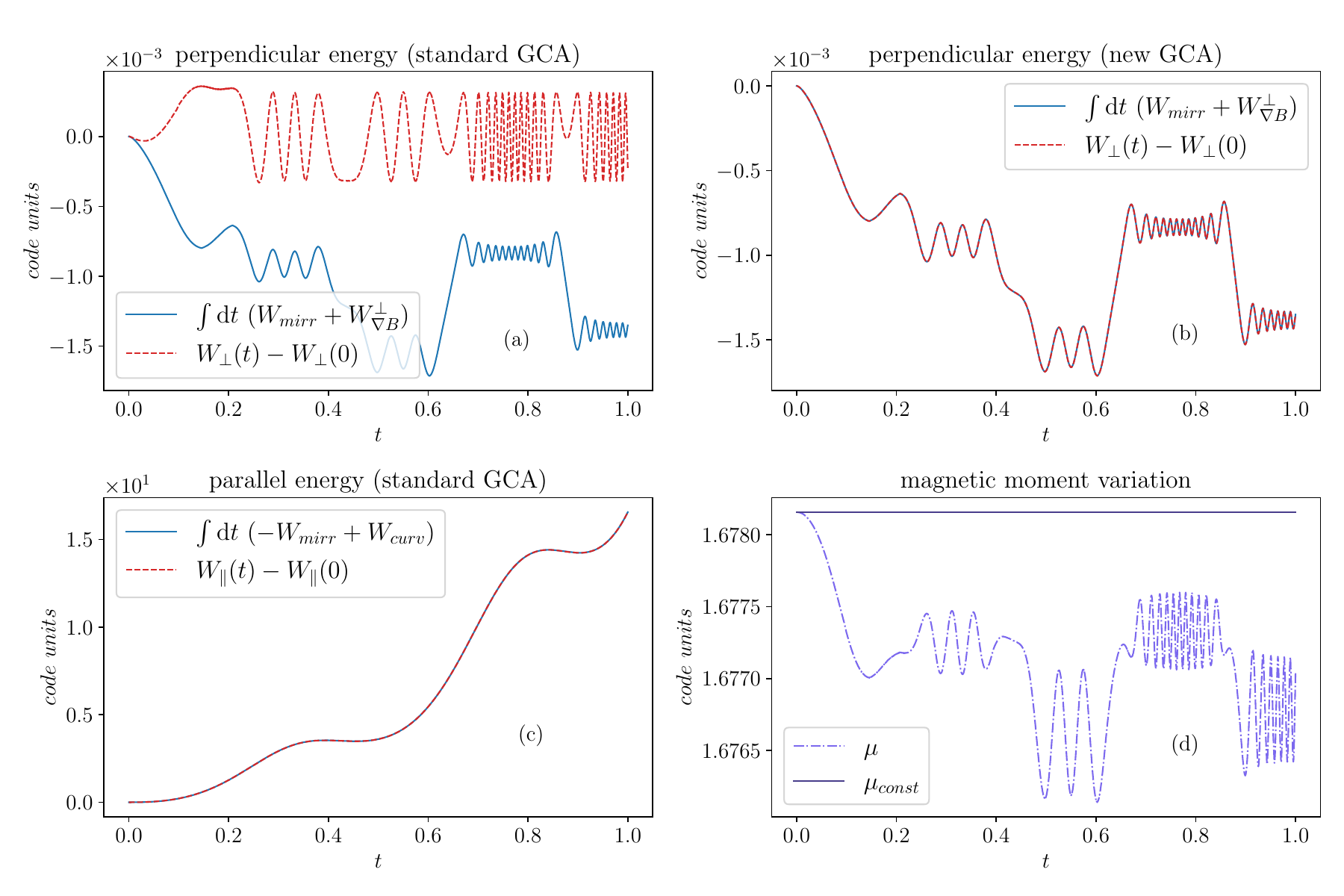}
 \caption{Left column: Failing perpendicular (a) and fulfilled parallel (c) energy conservation for the test particle evolving in the Alfvén-wave test case. Right column: (b) Fixed perpendicular energy after the new GCA implementation and (d) effect of the new equation on the magnetic moment evolution in time. $\mu_{const}$ is the constant value of the magnetic moment in the standard GCA evolution, while $\mu$ is the varying value in the current new energy-conserving GCA implementation.} \label{fig:AW}
 \end{figure*}
 
A distribution function, $f(W_i) \equiv \rmd N/\rmd W_i,$ inside the turbulent looptop at the initial time $t = 2.5t_0$ is shown in Fig.~\ref{fig:dNdWi} for each of these contributions. In this particular MHD setup, the strength of magnetic-mirror effects dominates over the three, followed by the perpendicular-gradient term. This aligns with the expectation that an environment with well-developed turbulence in the magnetic field has strong and significant gradient effects. However, given their nature, $W_{mirr}$ terms \eqref{eq:Wmirr} only transfer energy between parallel and perpendicular motion without producing net particle energisation. It is instead $W_{curv}$ and $W^\perp_{\nabla B}$ that contribute to the total particle-energy variation (see Eq.~\eqref{Wtot}). In our ideal-MHD runs (where parallel electric fields are absent by definition), it is expected that these effects manifest via continuous scattering of electrons across turbulent magnetic structures, promoting Fermi-type acceleration mechanisms (\citealt{Lemoine2022}).\\
A discrepancy arises, however, when ensembles of test particles are tracked individually during their evolution in the turbulent plasma using the standard GCA approach, as we show in the following sections.
The lack of energy conservation makes it impossible to infer any sensible observation of the energisation processes and of the role of different magnetic effects in the acceleration of particles. 
We now describe the relevant test case of a single Alfvén wave.

\subsection{Test case: Alfvén-wave propagation}\label{sec:AW}
For this test, we employed a 2.5D MHD setup ($(y,x)\in[0,10]\times[0,5]$) containing a single propagating Alfvén wave. Fluid velocity, $\bb{v}_{MHD}$, and magnetic-field, $\bb{B,}$ profiles were initialised with uniform $x$ components ($v_x =0$, $B_x=1$) and perturbed sinusoidal $y$ and $z$ components of the form $A\sin(k_1x  + k_2y)$ (with $A = 0.1$, $k_1 = 2\pi$, $k_2 = 2k_1$), and then rotated by an angle of $\alpha = \tan^{-1}(2)$ around the $z$-axis. The electric field, $\bb{E,}$ was initialised as $\bb{E} = - \bb{v}_{MHD} \btimes \bb{B}$. Fig.~\ref{fig:AW} shows the evolution of a test-particle with unitary mass and charge $ m = 1$ and $q = 1$ (cgs units), initialised with a random velocity of around $1$ (cgs units) at the centre of the 2D domain. \\
We then tracked the particle energy by taking the time integral form of Eqs~(\ref{Wpar}) and (\ref{Wperp}), and we obtain an evident difference between parallel and perpendicular energy evolution (panels (c) and (a), respectively). In both plots, dashed red curves represent the energy variation (perpendicular and parallel) with respect to the initial value (i.e. time integral of the left terms in Eqs.~(\ref{Wpar}-\ref{Wperp})), while light blue curves show the variation of the terms on the right of Eqs.~(\ref{Wpar}-\ref{Wperp}) integrated in time. The overlap of the two curves is what guarantees correct energy conservation, and while in the parallel direction it is satisfied, the standard GCA approach is herewith found to fail on energy conservation in the perpendicular direction. It is clear that the simple test case considered eliminates all possible errors coming from interpolation issues, and leaves us to conclude that standard GCA must be changed when the aim is to quantify energisation aspects in detail.
As we identified the same failure in the original turbulent looptop study, we can draw the following conclusions on the source of the non-conserved energy feature: \textit{(i)} it is independent of the type of evolving MHD background (e.g. trivial Alfvén wave or turbulent looptop); \textit{(ii)} it has to be related to the numerical treatment of the evolving system; and \textit{(iii)} it only relates to perpendicular dynamics, since in the parallel direction no errors are detected. 

In the derivation of the energy-conservation equations (\eqref{Wpar}--\eqref{Wperp}), we can observe how parallel and perpendicular motion are treated differently by the numerical algorithm. Parallel-energy conservation is explicitly solved in the standard GCA system (\ref{eq:gcadupardt}), while perpendicular-energy conservation is implicitly expressed by Eq.~\eqref{eq:gcadmudt} via magnetic-moment conservation, but not explicitly solved numerically. From a physical and analytic point of view (taking into account the appropriate approximations), Eqs.~\eqref{eq:gcadmudt} and \eqref{Wperp} are exactly the same, but numerically speaking this is not necessarily the case. Due to inevitable numerical approximations (e.g. interpolating quantities from the MHD grid to the particle position), it is not guaranteed that the code treats the two expressions the same way and that while solving the first, the second also remains fulfilled. For this reason, in this work we avoided energy inconsistency in the perpendicular direction by substituting Eq~\eqref{eq:gcadmudt} with Eq.~\eqref{Wperp} in the GCA system for particle motion. This way, the physics of particles remains the same, but both parallel and perpendicular energy conservation are solved explicitly. In practice, exact magnetic-moment conservation is sacrificed in favour of energy conservation; in this paradigm, it is the quantity $u_\perp^2=(\gamma v_\perp^2)$ that is evolved at every time step, and $\mu$ is only computed as a derived quantity, $\mu = mu^2_\perp/(2B/\kappa)$. Results for this method in the Alfvén-wave case are presented in Fig.~\ref{fig:AW} in the rightmost panels, (b) and (d). Panels (a) and (b) confront perpendicular-energy conservation before and after the new GCA implementation. In panel (d), we show the effect of integrating the new equation into the magnetic moment over time. The new approach proves efficient, and, despite no longer being exactly constant, $\mu$ only oscillates very mildly, with variations of the order of $10^{-4}$ around the initial value. All other features of particle motion (trajectories, encountered magnetic-field strength, etc.) remain unchanged with respect to the original standard approach.
While the loss of exact magnetic-moment conservation may appear concerning, we believe that this is a fair trade-off to eliminate energy errors. The latter may seriously compromise the analysis of particle energisation in a complex environment such as the turbulent looptop, particularly because of the strong mirror forces, which must be handled accurately to avoid spurious net energisation; this should only be driven by curvature and perpendicular-gradient terms. It is also reassuring to consider that $\mu$ is, by definition, only an approximate invariant of motion; therefore, limited variations in its value may be acceptable.

\section{Electron energisation in a turbulent coronal looptop}\label{sec:Elec}

\begin{figure*}
   \centering
    \includegraphics[width=17.5cm]{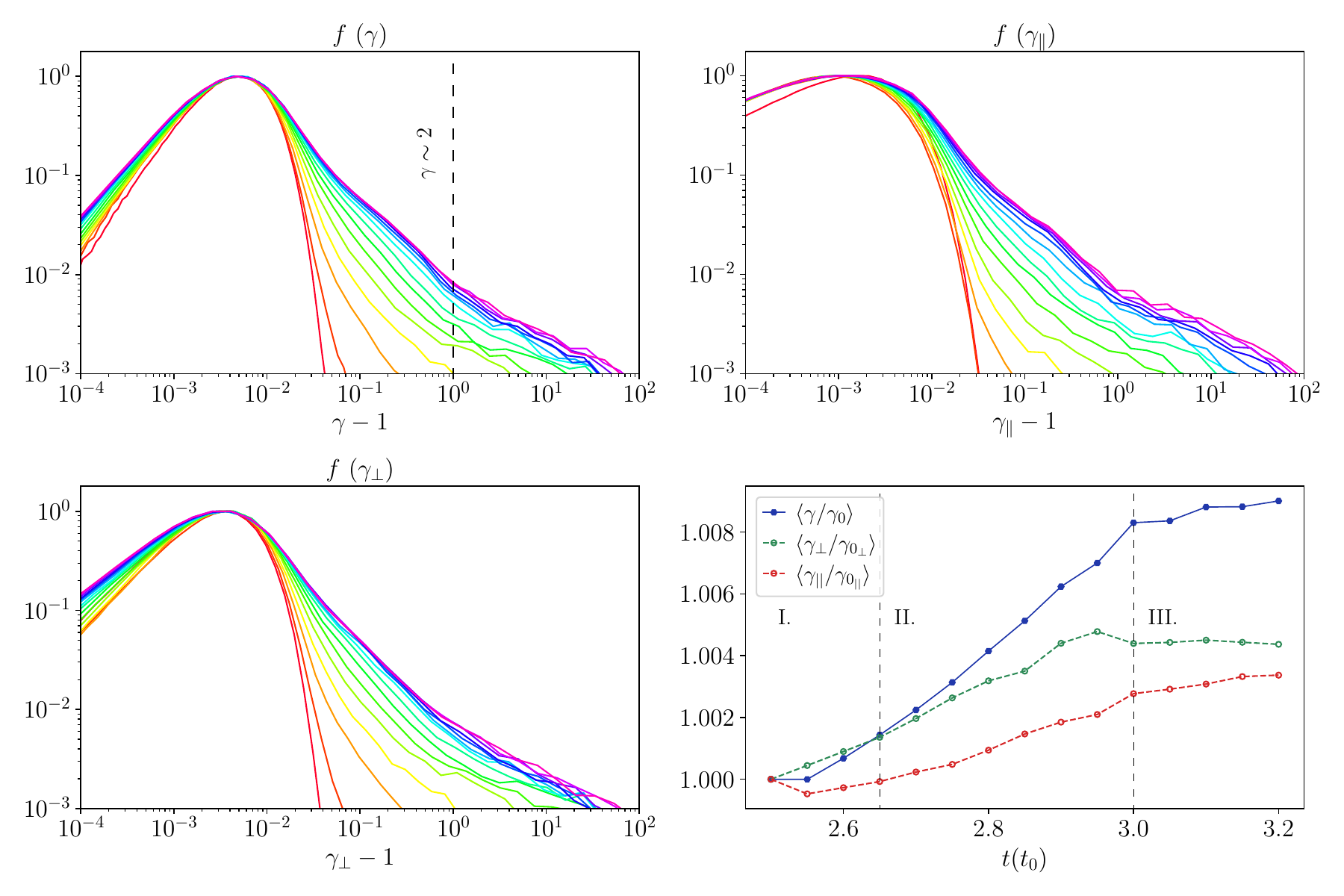}
      \caption{Evolution of electron distributions for total energy (top-left panel), parallel energy (top-right panel), and perpendicular energy (bottom-left panel) for $ t \in [2.5, 3.5]t_0$. Distributions are evenly spaced in time with a cadence of $\delta t = 0.05t_0$. The evolution of the mean energy relative to the initial energy, $\gamma/\gamma_0$ with $\gamma \leq 2$, is shown in the bottom right panel for the same time period.}
         \label{fig:edistrib}
   \end{figure*}

Building upon the improved numerical GCA framework established in the previous section, we now apply it to our electron
energy-distribution investigation and identify the relative importance of different mechanisms governing the acceleration process.
With the MHD setup described in Sect.~\ref{sec:numerical}, we performed test-particle simulations initialising $\sim 10^6$ electrons in the looptop region. The particle velocities were drawn from a Maxwellian distribution with temperature of $T = 20$~MK. We let particles evolve in the MHD background starting from the evolved MHD state at $t = 2.5t_0$ and running the simulation until the acceleration process settled into a steady-state condition. We recall that Fig.~\ref{fig:dNdWi} shows us the predominant role of gradient effects of the underlying MHD environment at initialisation time, and this aspect does not depend on the subsequent particle evolution. We follow this evolution up to reaching almost one MHD dynamical time unit at $t = 3.2t_0$ in co-evolving MHD settings.

In Fig.~\ref{fig:edistrib}, the time evolution of particle-energy distributions is shown, for a total $f(\gamma)\equiv \rmd N/\rmd\gamma$ (top-left panel), parallel $f(\gamma_\parallel)=f(\sqrt{1+u_\parallel^2/c^2})$ (top-right panel), and perpendicular $f(\gamma_\perp)=f(\sqrt{1+u_\perp^2/c^2})$ (bottom-left) energy. The distributions are plotted with a cadence of $\delta t= 0.05t_0$ for $ t \in [2.5, 3.2]t_0$. From the initial Maxwellian distribution (in red), we observe progressive energisation of electrons towards the high-energy range with the development of very extended suprathermal tails. \\
We observe that particles reach maximum energies of the order of $\gamma \sim 10^2$, which are rather unrealistic for the majority of electrons in the solar corona, even accounting for HXR-emitting populations (e.g. \citealt{Kruker2008, Lin_2003, Tomczak2001}).
We can explain this result by considering the various approximations employed in this work, which play a significant role in determining particle dynamics.\\
First, it is known that particle simulations in 2D geometries (with an ignorable third direction) are affected by numerical artefacts, potentially quenching cross-field-line
particle motion (e.g. \citealt{Jokipii, Jones_1998}). This allows for particles to be confined much longer than in realistic 3D geometries, where, instead, they would rapidly escape the system long before approaching such high energies. Electrons in our simulations are trapped and accelerated for periods of the order of one MHD timescale, $t_0=86$~s, which is much longer than their typical kinetic timescale of $2\pi/\omega_{pe} = 2\pi \sqrt{\epsilon_0 m_e/(n_e e^2)} \sim 3.5 \times 10^{-6}~$s (with $n_e = 10^9$~cm$^{-3}$ being the typical particle number density and $e=1.6 \times 10^{-19}$~C the particle charge, for this setup). Furthermore, assuming the test-particle approach is correct (i.e. that high-energy, non-thermal populations contain a negligible fraction of the looptop plasma energy), it remains to be evaluated how important kinetic effects discarded by the GCA paradigm could be.\\
For all these reasons, the following analysis focuses on an energy range only up to mildly relativistic energies of $\gamma \sim 2$ (dashed line in the top left panel of Fig.~\ref{fig:edistrib}), omitting the very-high-energy tails to maintain a more realistic connection with the observational energy range of HXR emission in the solar corona.

We can approximately identify three phases in the particle energy evolution; these are indicated (as I., II., and III.) in the bottom right panel of Fig.~\ref{fig:edistrib}, which shows the time evolution of the average variation in the total, parallel, and perpendicular energy with respect to the initial (at $t = 2.5t_0$) energy of each particle, $\gamma_0$:
\begin{itemize}
\item[I.] An injection phase from the initial time, $2.5t_0,$ until $2.65t_0$. Thermal electrons (initialised as described above) are injected in the pre-existing, turbulent MHD environment of the looptop and begin experiencing strong scatterings and interactions with electromagnetic fields. Hence, the particle ensemble rapidly adjusts to the underlying electromagnetic configuration, developing not necessarily physical features that only reflect this initial phase (e.g. a decrease in the average parallel energy and a mean perpendicular energy larger than the total one).

\item[II.] An acceleration phase from roughly $2.65t_0$ to $3.0t_0$. The system undergoes a steady energisation process.

\item[III.] A saturation phase, where the system's evolution slows down and the energy gain comes to a halt, introducing near-zero changes in the energy distributions.
\end{itemize}

\begin{figure*}
\centering
  \minipage{0.8\textwidth}
  \includegraphics[width=\linewidth]{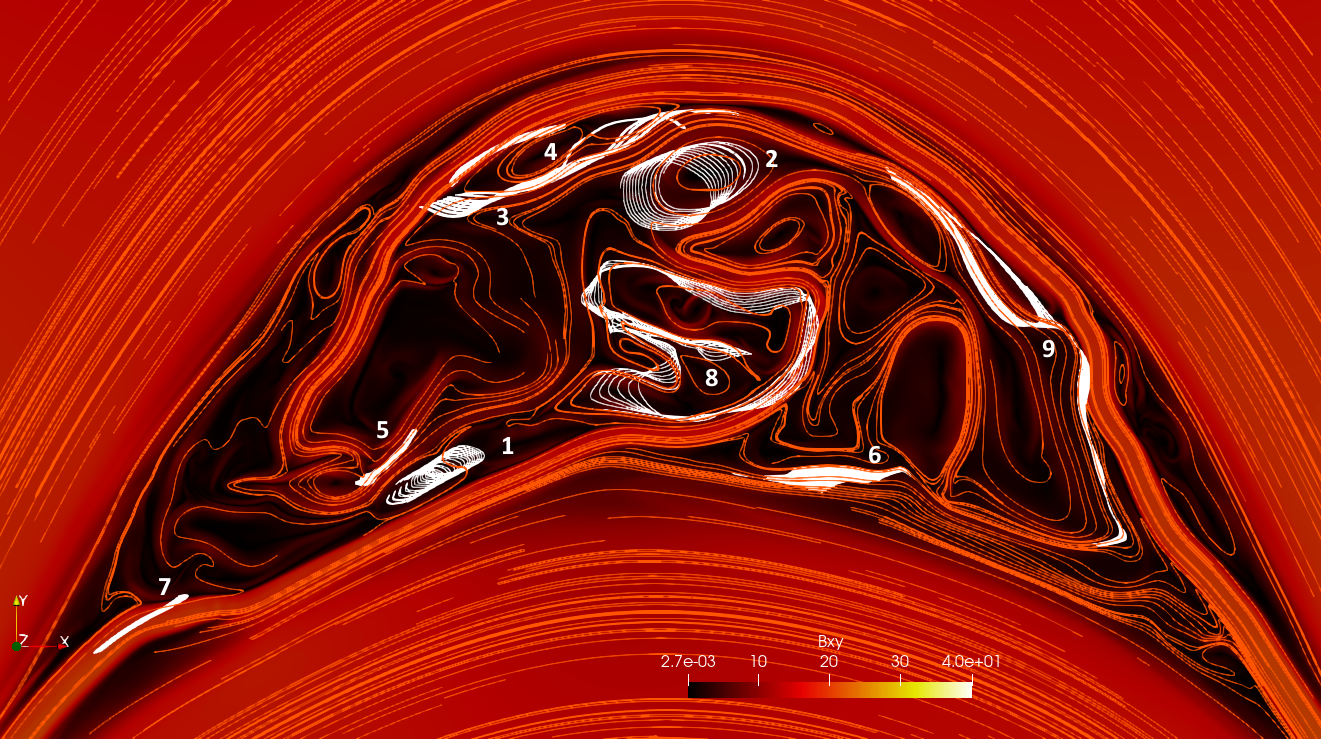}
 \endminipage \vspace{0.5cm} \\
 \minipage{\textwidth}
 \centering
      \includegraphics[width=0.95\linewidth]{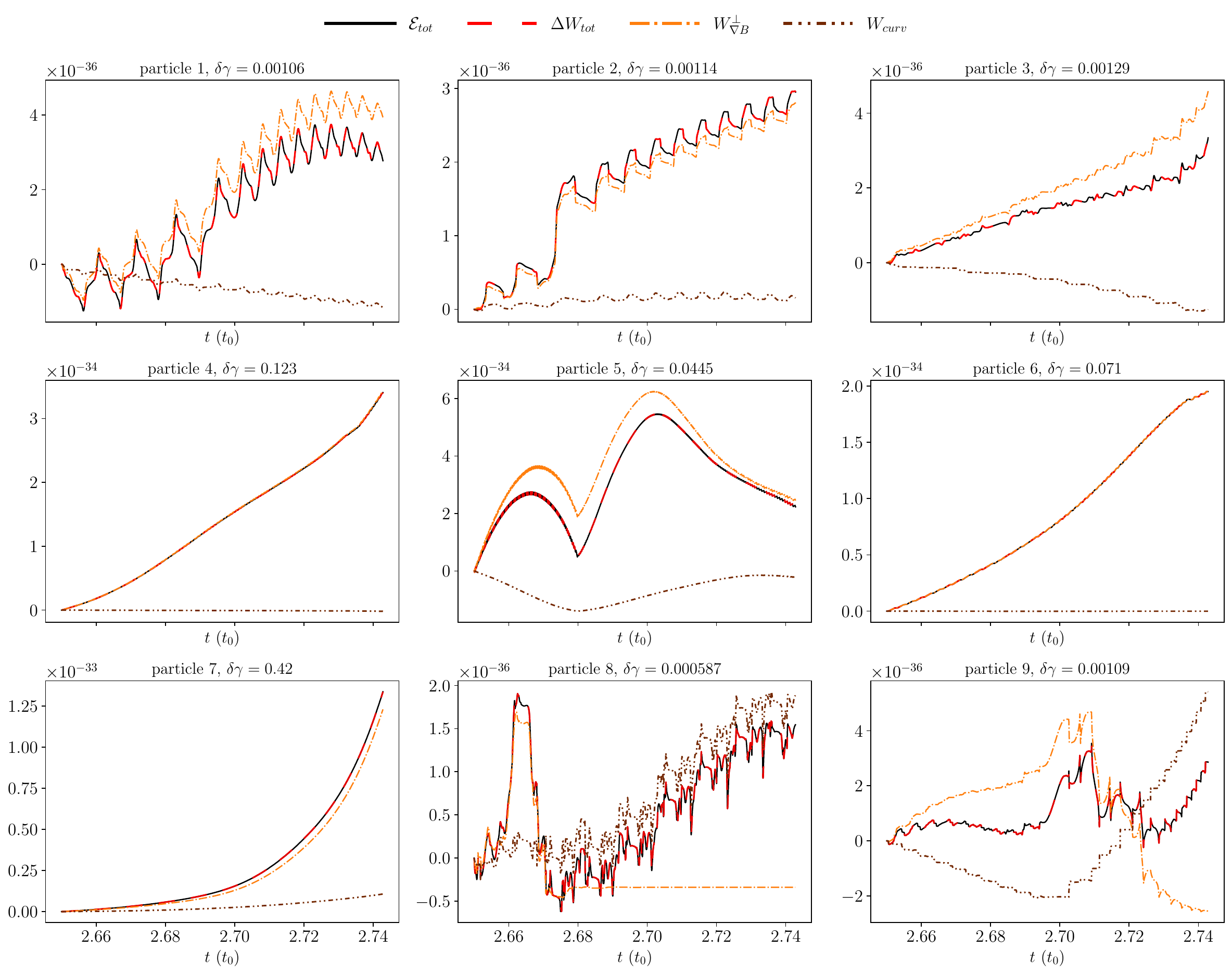}
 \endminipage 
 \caption{Top panel: trajectories of 9 representative particles during the time interval $[2.65-2.75 ] \ t_0$ on top of the 2D coronal loop magnetic field structure at $t = 2.65t_0$, coloured by the in-plane field intensity (code units) and threaded by field lines. Bottom: plots of the total energy evolution (integral form of Eq.~\eqref{Wtot}, in code units) decomposed into single-term contributions.}  \label{fig:trajandenergy}
 \end{figure*}
 
Overall, electrons in the initial population experience an average increase in $\gamma$ of $\sim 0.8\%$, which in this ideal-MHD setup is entirely attributed to Fermi-like processes (see e.g. \citealt{Guo_2019}, \citealt{Zhang_2021}, \citealt{Lemoine2022}). 
In the perpendicular direction, acceleration is slightly faster and stronger on average. This suggests the presence of efficient perpendicular-energisation mechanisms. \\
To investigate this further, we performed single-particle analyses on the total energy evolution of a fraction of electrons ($\sim 1000$) randomly sampled from the total ensemble. We track these electrons from $2.65t_0$ to $2.75t_0$, when the system has entered a steady state after the injection phase and acceleration is still strong, with a total energy increase of $\approx 0.2\%$ (bottom right panel in Fig.~\ref{fig:edistrib}). We report the outcome for nine representative particles in Fig.~\ref{fig:trajandenergy}, which shows how these particles experience different types of motions and magnitudes of acceleration. A similar behaviour is then observed in the remaining population.

The upper panel of Fig.~\ref{fig:trajandenergy} is a visualization of the 2D coronal-loop magnetic-field structure at $t=2.65t_0$, coloured by the in-plane magnetic-field intensity and overlaid with selected field lines. Nine particle trajectories are shown in white. In the bottom panels, we show the total energy variation for each particle, $\delta\gamma=\gamma(t=2.75t_0)-\gamma(t=2.65t_0)$, together with the contribution of the individual energisation terms introduced in Sect. \ref{sec:sec3_1}. Specifically, $\mathcal{E}_{tot} = W_{curv} + W^\perp_{\boldsymbol{\nabla} B}$ is the integral in time of the right side of Eq.~\eqref{Wtot}, and $\Delta W_{tot}$ is the integral of the left side. Energy conservation is achieved as indicated by the overlap of $E_{tot} $ and $ \Delta W_{tot}$. For all particles, the fluctuating behaviour of the total energy change suggests that \textit{\emph{stochastic acceleration}} may be at play (e.g. \citealt{Bian_2012}). This analysis also provides insight into the role of the different acceleration terms: there is a net dominance of perpendicular-gradient effects, $W^\perp_{\boldsymbol{\nabla} B}$, in the contribution to total energy increase, which agrees with the dominant perpendicular-energy increase previously described in the energy-distribution analysis. Curvature effects, $W_{curv}$, appear to play a more significant role for particles in closed trajectories around magnetic islands or exploring larger areas of the domain (e.g. particles 3, 8, and 9 in Fig.~ \ref{fig:trajandenergy}), although they do not drive very strong energisation. Indeed, the largest $\delta \gamma$ is associated with fast bouncing motions confined within small regions of space (e.g. particles 4, 6, and 7).

\begin{figure}
    \includegraphics[width=\linewidth]{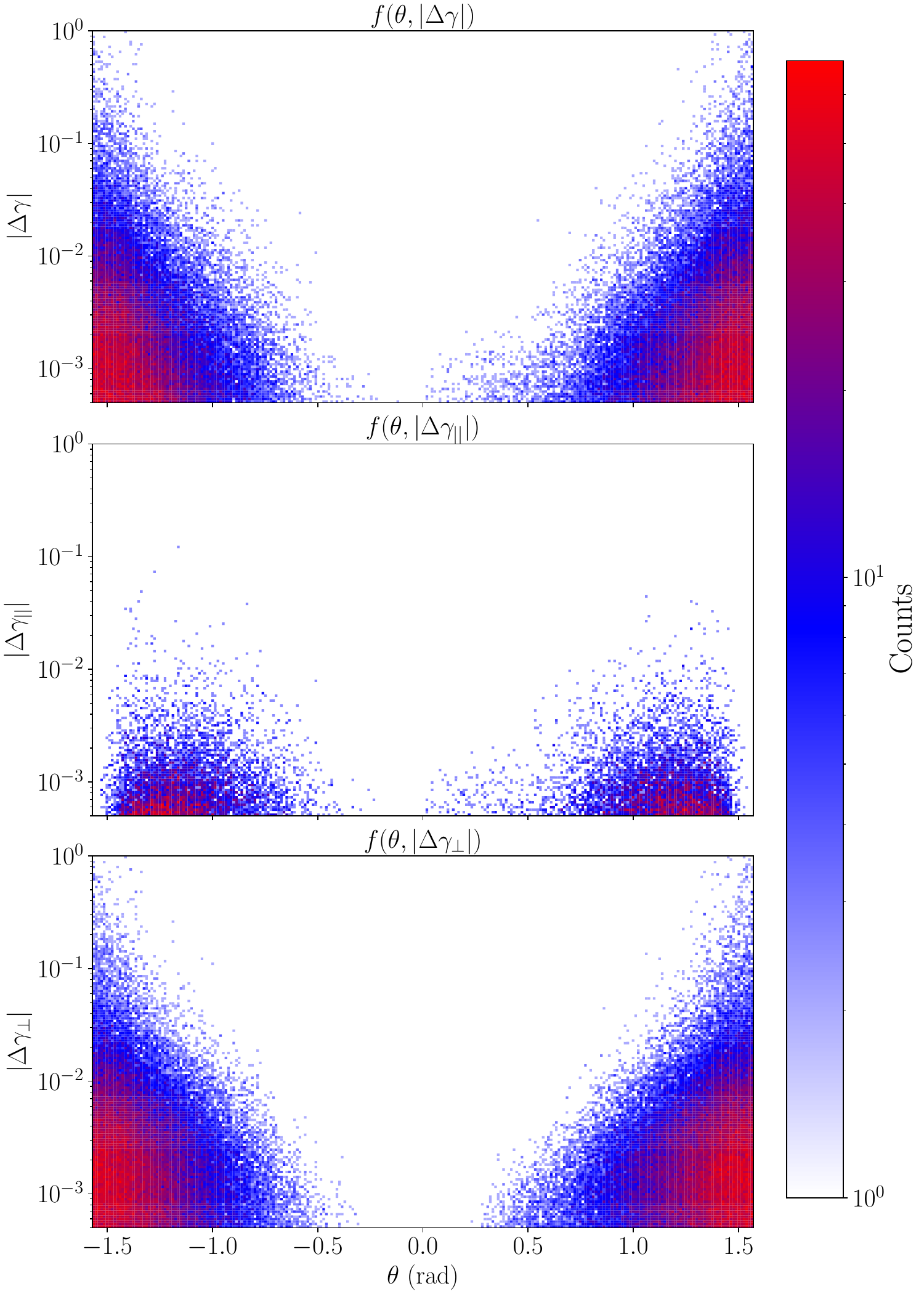}
    \caption{2D distributions at $t=2.75t_0$ of particle energy gain, $\Delta \gamma = \gamma(t_2 =2.75t_0) - \gamma(t_1=2.65t_0),$ as a function of pitch angle, $\theta,$ for total, parallel, and perpendicular energies.}
    \label{fig:stat}
\end{figure}

As also discussed in \citealt{Bacchini2024} (which we hereby augment with details on the particle energetics, as the original study was affected by non-conserved energetics of standard GCA), many particles experience long-term confinement between magnetic islands, which influences their energy gain. We further explored the connection between energy gain and type of particle trajectory with the statistical analysis shown in Fig.~\ref{fig:stat}. The figure shows, from top to bottom, 2D distributions at $t=2.75t_0$ of the total, parallel, and perpendicular energy gain, $\Delta \gamma$, as a function of the particle pitch angle, $\theta=\tan^{-1}(v_\perp/v_\parallel)$ 
(where $v_\parallel$ is the particle's velocity component along the magnetic-field line and $v_\perp$ is the component perpendicular to the guiding-centre motion). The pitch angle was used as an estimate of the degree of confinement for particles; indeed, its value determines the necessary condition for particles to be trapped in a magnetic-mirror configuration. Hence, large pitch angles indicate the likelihood of the particle to experience a bouncing motion. The statistical ensemble was chosen from the global $\sim 10^6$ evolving particles by selecting the population of test electrons that evolve in the time interval $[t_1=2.65t_0,t_2=2.75t_0]$ with a pitch angle such that $\theta_{t_2} \geq 0.9 \ \theta_{t_1}$. This filter was imposed to ensure that particles that are moving along magnetically confined trajectories maintain a pitch angle large enough to remain trapped for the whole period. The energy gain is computed as $\Delta \gamma=\gamma_{t_2} - \gamma_{t_1}$. From this analysis, we observe that the largest $\Delta \gamma$ are associated with the largest pitch angles, thereby confirming that the majority of electrons experiencing strong energisation are those trapped in bouncing trajectories and under the action of strong perpendicular-gradient effects. Furthermore, we observe again how the evolution and increase of total energy are dominated by the perpendicular dynamics, and parallel acceleration is weaker. 

These results allow us to firmly establish that the driver of strong particle energisation in a turbulent coronal-loop environment is perpendicular acceleration due to perpendicular-gradient effects. In the next section, we discuss our results in more detail.

\section{Conclusions}\label{conclusions}
Throughout this work, we revisited test-particle simulations of electrons evolving in a turbulent post-flare coronal loop, with the aim of characterising the aspects of their energisation in a 2.5D, ideal-MHD setup in MPI-AMRVAC 3.0 (\citealt{keppens2023}). We started from the approach and results presented in \citealt{Bacchini2024} and improved and extended their analysis specifically to obtain a precise measurement of particle energisation.

First, we established a consistent numerical framework to track particle energetics by implementing an improved guiding-centre-approximation formalism, which enforces explicit energy conservation while retaining the essential physics of particle dynamics. The new method evolves the perpendicular energy explicitly, and sacrifices exact magnetic-moment conservation to capture particle energisation much more accurately compared to standard approaches.
We observed how, in our turbulent MHD setup, magnetic-mirror terms dominate in instantaneous strength but, as expected, only act in exchanging energy between degrees of freedom. The net energisation arises from curvature and perpendicular gradient effects, which together account for an overall average electron acceleration of $\sim 0.8\%$ the initial energy, over one dynamical MHD time unit. Our improved numerical approach captures this dynamics accurately, without allowing mirror forces to introduce spurious net energisation. 
We observed that particles in the looptop turbulence develop supra-thermal tails in their energy distributions, with a relative increase in energy that is slightly more pronounced in the perpendicular direction. Although the very-high-energy cut-offs observed ($\gamma\sim10^2$) are likely an artefact of the approximations introduced in our approach, the moderate-energy parts of the distributions are consistent with flare observations in the solar corona. The bulk of the energy distribution remains within the range of SXR (energy $\lesssim 10$~keV and $\gamma \lesssim 1.02$), while the non-thermal tails reach energies compatible with HXR Bremsstrahlung emission (energy $ > 10$~keV, $\gamma \sim [1.02-2]$).\\
We distinguished three stages in the evolution: I. an initial injection phase marked by an adjustment of the initially thermal electrons to the turbulent environment; II. a sustained acceleration phase characterised by efficient energy gain; and III. a saturation phase with slower evolution and converged, unchanging non-thermal features.\\
The analysis of energy-exchange terms indicates that perpendicular-gradient effects dominate the acceleration and efficiently pump energy into the particle motion, while curvature effects play a secondary role, mainly for particles travelling on extended trajectories. Statistical correlations between energy gain and pitch angle further confirm that the strongest energisation occurs for electrons trapped in bouncing trajectories, which is consistent with stochastic, Fermi-like acceleration mechanisms mediated by turbulence. This is in turn consistent with the analysis of \citealt{Bacchini2024}, supported here by quantitative measurements allowed by our new numerical approach. \\
Overall, these results support the interpretation that KHI-induced looptop turbulence provides an efficient mechanism to produce non-thermal electrons in post-flare coronal loops, thereby contributing to the observed X-ray emission. Although the study here adopted an idealised 2.5D MHD background that does not incorporate the reconnection zone above the post-flare loop, our findings are expected to carry over to turbulent flare arcades as found in recent full-3D MHD settings of the standard flare evolution \citep{Ruan_2023,Ruan2024}. 

Although our 2D test-particle approach may overestimate confinement time and maximum energies, neglecting feedback and kinetic effects and requiring more quantitative validation, it nonetheless offers a valuable new framework to address the multistage process of magnetic-to-particle energy conversion in flare modelling. We can also use this to study proton-acceleration aspects in the more realistic 3D setups. More advanced numerical approaches, for example\ combining the GCA with the full equations of motion (\citealt{bacchini2020}), could also be enhanced with our improved GCA formalism and applied to the study presented here. \\
The standard GCA is perfectly suitable for cases where particles do not experience strong net energisation; i.e. their total energy is roughly constant, which is the case when the dynamics is strongly dominated by magnetic fields (e.g. mainly with mirroring effects), and directly accelerating electric fields are absent or negligible. The energy-conserving formulation can be adopted when particle energisation is the main focus of the study and electric fields have a relevant effect on the long-term evolution.\\
The modifications to the standard GCA algorithm presented here could be highly relevant in other, not purely solar, settings and are expected to lead to more physically sound findings on particle energisation in turbulent plasmas. The new energy-conserving GCA treatment allows one to directly quantify any resulting deviations from the first adiabatic invariance (constant magnetic moment). Our demonstration of its effect on a simple Alfv\'en-wave setup can be used to check independent implementations of particle GCA integrators.\\

\begin{acknowledgements}
RK and FB acknowledge funding from the KU Leuven C1 project C16/24/010 UnderRadioSun. RK also acknowledges the Research Foundation Flanders FWO project G0B9923N Helioskill.
FB also acknowledges support from the FED-tWIN programme (profile Prf-2020-004, project ``ENERGY'') issued by BELSPO, and from the FWO Junior Research Project G020224N granted by the Research Foundation -- Flanders (FWO). CM acknowledges support from the FWO PhD Fellowship fundamental research 1105426N granted by the Research Foundation -- Flanders (FWO).
\end{acknowledgements}

\bibliographystyle{aa} 
\bibliography{bib} 

\end{document}